# Chemical Pressure Tuning of Multipolar and Magnetic Orders in Ba$_2$(Cd$_{1-x}$Ca$_x$)ReO$_6$ Double Perovskites

Koki Shibuya[1], Daigorou Hirai[1], and Koshi Takenaka[1]

[1]Department of Applied Physics, Nagoya University, Nagoya 464–8603, Japan

**ABSTRACT**. Double perovskite compounds containing $5d$ transition metal elements have been extensively studied as platforms for multipolar order phenomena stemming from spin-orbit-entangled $5d$ electrons. In this study, we examine the interplay between crystal structure, multipolar order, and magnetic order in solid solutions of double perovskites with the $5d^1$ electronic configuration: Ba$_2$CdReO$_6$ and Ba$_2$CaReO$_6$, which exhibit distinct electronic orders. The substitution of larger Ca$^{2+}$ ions for Cd$^{2+}$ in Ba$_2$CdReO$_6$, systematically increases the lattice constant with increasing the amount of substitution $x$. Although the spin-orbit-entangled $J = 3/2$ state remains intact upon substitution, both the quadrupolar order below $T_q$ = 25 K and the canted antiferromagnetic (AFM) order below $T_m$ = 12 K in Ba$_2$CdReO$_6$ are progressively suppressed as $x$ increases. Magnetization measurements reveal that the canted AFM order is suppressed at $x$ = 0.6, transitioning to a colinear AFM order, while the quadrupolar order persists up to $x$ = 0.9. The experimental electronic phase diagram, summarizing the dependence of electronic orders on lattice constants, aligns well with the theoretical phase diagram considering electric quadrupolar interactions [G. Chen *et al*., Phys. Rev. B **82**, 174440 (2010)]. This correspondence confirms that chemical pressure induced by substitution effectively tunes the interaction between $5d$ electrons. The results highlight the potential of chemical pressure to modulate multipolar interactions, paving the way for novel multipolar properties in $5d$ electron systems.

## I. INTRODUCTION

In $5d$ transition metal (TM) compounds, strong spin-orbit interactions (SOIs) and electron correlations result in the formation of electronic states characterized by the total angular momentum $J$, where spin and orbital degrees of freedom are coupled [1]. These states underpin many novel electronic phases, including the Kitaev spin liquid and correlated topological semimetal [2] [3] [4] [5]. Among these, multipolar order has emerged as a particularly intriguing phenomenon. For instance, in the double perovskite oxide Ba$_2$MgReO$_6$, the $5d$ electrons in Re$^{6+}$ ions form a four-fold degenerate $J = 3/2$ state, which leads to the manifestation of electric quadrupole order [6] [7] [8] [9]. Similarly, in Ba$_2$BOsO$_6$ ($B$ = Zn, Mg, Ca), $5d$ electrons in Os$^{6+}$ ions are reported to exhibit magnetic octupolar order [10] [11]. Thus, double perovskites represent an essential platform for studying multipolar orders in $5d$ systems [12].

While multipolar orders in $4f$-electron compounds have been extensively studied [13] [14] [15], their counterparts in $5d$ systems are expected to exhibit distinct behavior due to differences in interaction strengths and electronic states. Theoretical studies have predicted various multipolar orders in double perovskites with $d^1$ electronic configurations [4] [16]. These predictions incorporate the electric quadrupolar interaction ($V$) along with nearest-neighbor antiferromagnetic ($J$) and ferromagnetic ($J'$) interactions. The relative strengths of these interactions define phase boundaries, enabling the realization of diverse electronic orders, such as collinear antiferromagnetic (AFM) order with a large staggered octupole moment.

Experimental studies have demonstrated that these interactions can be modified through the application of pressure. For instance, applying pressure to Ba$_2$MgReO$_6$ results in the disappearance of the quadrupolar-ordered phase at approximately 5 GPa, coinciding with a transition from canted AFM (cAFM) to collinear AFM order [17]. This shift is attributed to changes in the relative strengths of $J$, $J'$, and $V$ due to pressure-induced reductions in TM-TM and TM-O distances. However, theoretical analyses suggest that isotropic pressure alone induces only minor changes in these interactions and that anisotropic pressure may be required for significant phase transitions [18]. Limitations in experimental techniques under pressure further motivate alternative approaches to modulate electronic phases.

In this context, we employed chemical substitution as a means to simulate "chemical pressure" by altering the ionic radii of constituent ions. Double perovskite compounds with the formula $A_2BB'$O$_6$ allow the tuning of TM-TM and TM-O distances through substitution at the $B$ site [Fig. 1(a)]. This approach facilitates precise control of interaction strengths and enables a detailed investigation of electronic phase boundaries.

Our study focuses on double perovskites Ba$_2B$ReO$_6$ ($B$ = Mg, Zn, Cd, and Ca), where Re$^{6+}$ ions with $d^1$ electronic configurations occupy the $B'$ site, and nonmagnetic Ba$^{2+}$ ions occupy the $A$ site. When the $B$ site is populated with divalent cations of varying ionic radii [Mg$^{2+}$ (0.720 Å) < Zn$^{2+}$ (0.740 Å) < Cd$^{2+}$ (0.95 Å) < Ca$^{2+}$ (1.00 Å)] [19], the lattice constants of the resulting Ba$_2B$ReO$_6$ compounds vary systematically (Ba$_2$MgReO$_6$ ($a$ = 8.0849(2) Å) < Ba$_2$ZnReO$_6$ ($a$ = 8.1148(1) Å) < Ba$_2$CdReO$_6$ ($a$ = 8.3253(1) Å) < Ba$_2$CaReO$_6$ ($a$ = 8.37446(4) Å)) [6] [7] [20] [21] [22] [23] [24] [25] [26] [27]. This trend directly impacts the interaction strengths and, consequently, the electronic phases.

Previous studies have revealed that Ba$_2$MgReO$_6$ exhibits cAFM and quadrupolar order at $T_m$ = 18 K and $T_q$ = 33 K, respectively [7] [8] [9] [20] [28] [29]. In Ba$_2$ZnReO$_6$ and Ba$_2$CdReO$_6$, cAFM order persists, accompanied by spontaneous magnetization below $T_m$ = 11 K and 12 K, respectively [6] [23] [30]. These findings suggest a shared magnetic ground state across these compounds. In contrast, Ba$_2$CaReO$_6$, with the largest lattice constant, exhibits only collinear AFM order at $T_m$ = 16 K and lacks any indication of quadrupolar order [26]. This discrepancy suggests the existence of an electronic phase boundary between Ba$_2$CdReO$_6$ and Ba$_2$CaReO$_6$, highlighting the potential of chemical pressure to drive phase transitions.

In this study, we synthesized solid solutions of $Ba_2CdReO_6$ and $Ba_2CaReO_6$, denoted as $Ba_2(Cd_{1-x}Ca_x)ReO_6$ ($x$ = 0–1), to investigate the effects of chemical pressure on electronic phases. The results demonstrate that chemical pressure effectively tunes interaction strengths, suppressing cAFM and quadrupolar order. The experimentally derived phase diagram confirms the efficacy of chemical substitution as a tool to explore multipolar interactions in 5$d$ systems.

## II. EXPERIMENTAL

Polycrystalline samples of $Ba_2(Cd_{1-x}Ca_x)ReO_6$ ($x$ = 0, 0.25, 0.3, 0.4, 0.5, 0.55, 0.6, 0.75, 0.8, 0.9, 1) and $Ba_2CdWO_6$ were synthesized via conventional solid-state reactions. Stoichiometric amounts of BaO, CaO, CdO and $ReO_3$ (or $WO_3$) powders were thoroughly mixed and pelletized in a nitrogen-filled glove box. The pellet was placed in an alumina tube (inside a gold tube only for $x$ = 1) and sealed within an evacuated quartz tube. The initial heating was conducted for 24 hours at 700°C for $x$ = 0 and $Ba_2CdWO_6$, for 32 hours at 800–850°C for $x$ = 0.25–0.9, and for 48 hours at 750°C for $x$ = 1. The resulting sample was then ground and repelletized before undergoing a second heating cycle under identical sealed conditions. This step involved heating for 24 hours at 700°C for $x$ = 0, for 32 hours at 800–850°C for $x$ = 0.25–0.9, for 48 hours at 900°C for $x$ = 1, and for 24 hours at 800°C for $Ba_2CdWO_6$. To enhance the degree of cation ordering at the $B'$ site ($Re^{6+}$) and $B$ site ($Ca^{2+}$ and $Cd^{2+}$), solid-solution samples ($x$ = 0.25–0.9) were annealed at relatively low temperatures of 650–750°C for 96 hours in an evacuated quartz tube.

The synthesized samples were characterized using x-ray diffraction (XRD) measurements at room temperature with Cu-$K\alpha$ radiation on a diffractometer (MiniFlex; Rigaku). Lattice constants were determined by fitting the XRD patterns with SmartLab Studio II software (Rigaku).

Linear thermal expansion $\Delta L(T)/L$ was measured using a strain gauge (KFL; Kyowa Electronic Instruments Co. Ltd.) and a copper reference [31]. Magnetization measurements were performed using a Magnetic Property Measurement System (MPMS3; Quantum Design), with the samples mounted on a quartz holder and secured with varnish (GE7031; General Electric Company). Heat capacity measurements were conducted via the relaxation method using a physical properties measurement system (PPMS; Quantum Design). The polycrystalline samples were mounted on a sapphire sample stage with grease (Apiezon-N; Materials Ltd.).

## III. RESULTS
### A Structural change

All peaks in the powder XRD pattern of the synthesized $Ba_2(Cd_{1-x}Ca_x)ReO_6$ ($x$ = 0–1) samples (Fig. S1 in Supplementary Material) were successfully indexed, assuming a cubic double-perovskite-type structure (space group $Fm\bar{3}m$) without impurities. This confirms that all polycrystalline $Ba_2(Cd_{1-x}Ca_x)ReO_6$ samples exhibit undistorted double-perovskite-type structures at room temperature, consistent with previous reports for $Ba_2CdReO_6$ and $Ba_2CaReO_6$.

The 6 4 2 reflections of $Ba_2(Cd_{1-x}Ca_x)ReO_6$ samples shift systematically to lower angles with increasing Ca content $x$ (Fig. S1 in Supplemental Material), indicating an increase in lattice constant. Figure 1(c) presents the compositional dependence of the lattice constant obtained from XRD pattern fittings. The lattice constants of the end members ($x$ = 0 and 1) align well with the previous reports [23] [26], and the lattice constants of the solid solutions increase approximately linearly with $x$, consistent with Vegard's law. Thus, double perovskites with systematically controlled lattice constants and Re-Re distances, i.e., chemical pressure, were successfully synthesized.

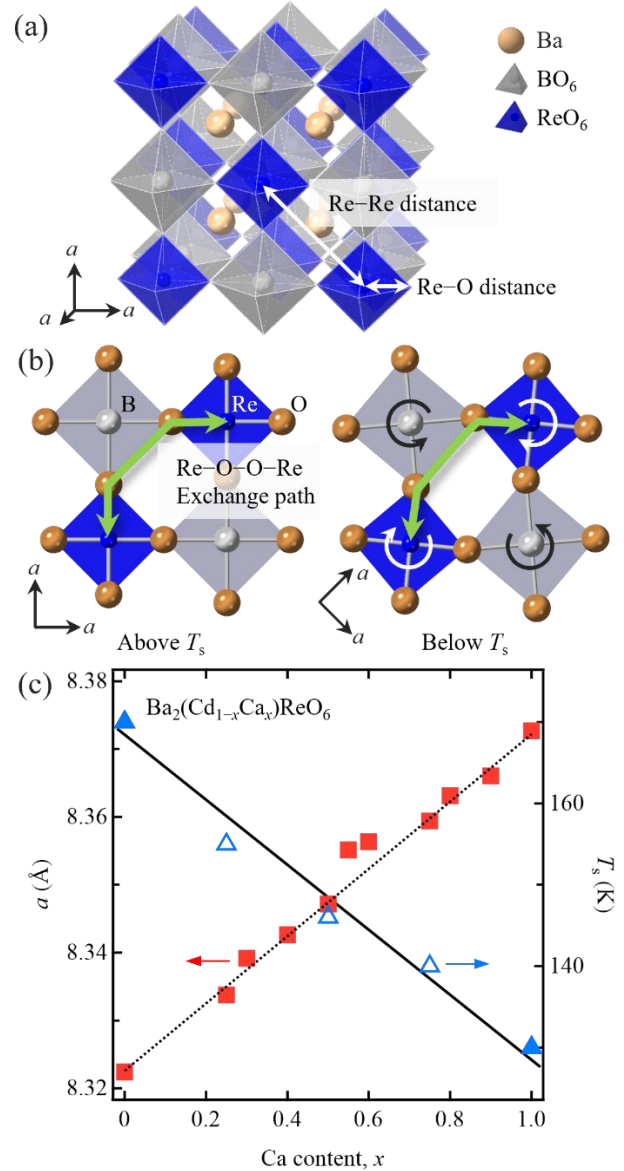

FIG. 1 (a) Crystal structure of the cubic double perovskite $Ba_2BReO_6$. (b) Re–O–O–Re exchange path above and below the cubic-to-tetragonal structural transition. (c) Compositional dependences of the lattice constant, $a$, for $Ba_2(Cd_{1-x}Ca_x)ReO_6$ ($x$ = 0–1), determined from XRD patterns at room temperature (red filled squares, left axis), and structural phase transition temperatures, $T_s$, obtained via linear thermal expansion measurements (blue open triangles, right axis). $T_s$ values for $x$ = 0 and 1 (blue filled triangles) are taken from references [23,25].

The phase stability of the cubic double perovskite is quantified using the tolerance factor [32]. The tolerance factors of $Ba_2CdReO_6$ and $Ba_2CaReO_6$ are 0.990 and 0.979, respectively [23], indicating structural instability. Both compounds exhibit structural phase transitions from cubic (space group $Fm\bar{3}m$) to tetragonal (space group $I4/m$) at $T_s$ = 170 and 120 K, respectively [23] [25]. Such structural phase transitions are commonly observed in other double perovskites and arise from the out-of-phase rotations of octahedra about the $c$-axis to optimize the B–O bonding [Fig. 1(b)] [12]. As a consequence of the structural phase transition, the deformation of $ReO_6$ remains minimal, at less than 1%. However, the Re–O–O bond angles in the exchange path, which are 135° in the cubic phase, undergo a significant transformation to approximately 125° and 145° [27]. These transitions manifest as kinks in the linear thermal expansion [23]. For the solid solutions, the structural phase transition temperature, $T_s$, were determined via linear thermal expansion measurements. A kink in linear thermal expansion was observed between 170 and 120 K for all measured solid solutions (Fig. S2). $T_s$ decreases linearly with $x$ [Fig. 1(c)], confirming that Ca substitution systematically applies chemical pressure.

**B Magnetism**

Figure 2(a) shows the temperature dependence of magnetic susceptibility for $Ba_2(Cd_{1-x}Ca_x)ReO_6$ ($x$ = 0.25–0.9) under an applied magnetic field of 0.1 T. Samples with $x$ = 0.25–0.55 display a sharp increase below 10 K, characteristic of a cAFM transition. In contrast, samples with $x$ = 0.6–0.9 exhibit a gradual increase, suggesting a transition of magnetic ground state between $x$ = 0.55 and 0.6. The field dependence of magnetization at 2 K [Fig. 2(b)] supports this, showing hysteresis and spontaneous magnetization for $x$ = 0.25–0.55, while $x$ = 0.6–0.9 shows linear field dependence without hysteresis. This implies cAFM order for $x \leq 0.55$ and collinear AFM order for $x \geq 0.6$.

The saturation magnetic moment, $M_{sat}$, was determined by extrapolating magnetization data between 5 and 7 T to $B$ = 0. $M_{sat}$ decreases gradually with increasing $x$ and vanishes above $x$ = 0.6 [Fig. 2(b), inset]. At $x$ = 0, $M_{sat}$ is $0.16\mu_B$, consistent with previous reports [23]. Although smaller than typical ferromagnetic values, $M_{sat}$ is significantly larger than that of standard canted antiferromagnets. This enhancement arises from an unusually large canting angle stabilized by high-temperature quadrupole order, as discussed later. The reduction in $M_{sat}$ with $x$ reflects a decreasing cant angle, with the cant angle disappearing at $x$ = 0.6, marking a magnetic state transition from cAFM to collinear AFM.

The magnetic transition temperature ($T_m$) for $x$ = 0–0.55 (cAFM phase) was identified as the temperature at which a sharp rise in the magnetic susceptibility occurs. Although the magnetic ground state for $x$ = 0.6–1 appears to be collinear AFM order, a kink characteristic of collinear AFM transitions was not observed. Hence, $T_m$ was defined as the temperature corresponding to a peak in heat capacity, as shown in the inset of Fig. 2(a). This peak aligns approximately with the deviation in the inverse susceptibility at 7 T from a straight line below $T_q$. The lack of a visible kink at the magnetic transition is likely due to B-site disorder caused by random Ca and Cd occupation and geometric frustration discussed later. At $x$ = 0.25, where the cAFM phase transition occurs, a distinct heat capacity peak is observed at $T_m$. Conversely, for $x$ = 0.9, the heat capacity gradually increases above $T_m$ with a smaller peak at $T_m$ [Fig. 2(a) inset], despite better crystallinity compared to $x$ = 0.25 based on powder XRD patterns. This behavior is characteristic of frustrated magnets, where short-range correlations develop at temperatures higher than $T_m$. The value of $T_m$ decreases slowly with increasing $x$ until a sharp drop occurs as the cAFM phase transitions to collinear AFM. Within the collinear AFM phase, $T_m$ becomes nearly independent of $x$.

Figure 3(a) presents the compositional dependence of the effective magnetic moment ($\mu_{eff}$) and Weiss temperature ($\Theta_W$) for $Ba_2(Cd_{1-x}Ca_x)ReO_6$ ($x$ = 0–1), derived from Curie-Weiss fitting of inverse magnetic susceptibility at 7 T between 100 and 350 K. The effective magnetic moments and Weiss temperatures for $x$ = 0 and 1 are estimated as $0.68(4)\mu_B$, −13.2(2) K, and $0.66(4)\mu_B$, −29.1(3) K, respectively, consistent with previous studies [23] [26]. The effective magnetic moment remains nearly constant at approximately $0.67\mu_B$ for all compositions, whereas the Weiss temperature decreases gradually until $x$ = 0.6 and then significantly for $x$ > 0.6.

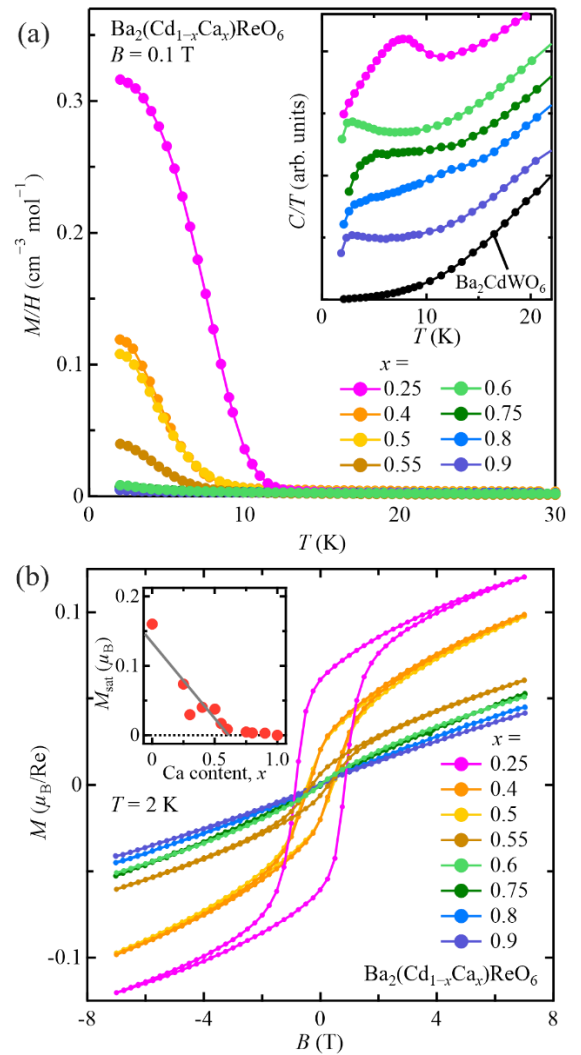

FIG. 2. (a) Temperature dependence of the magnetic susceptibilities under a 0.1 T applied magnetic field, and (b) field dependence of magnetization at 2 K for $Ba_2(Cd_{1-x}Ca_x)ReO_6$ ($x$ = 0.25–0.9). The inset in (a) shows the temperature dependence

of the heat capacity near the magnetic transition temperatures for $x = 0.25, 0.6, 0.75, 0.8, 0.9$, and $Ba_2CdWO_6$. The inset in (b) illustrates the compositional dependence of saturation moments ($M_{sat}$) for $x = 0$–1. The gray solid line serves as a guide for the eye.

The observed $\mu_{eff}$, considerably smaller than the expected $1.73\mu_B$ for spin-1/2, arises from the cancellation of spin and orbital angular momentum. Similar reductions in effective magnetic moments are noted in other $5d^1$ double perovskites, such as $Ba_2MgReO_6$ ($0.68\mu_B$) [7] and $Ba_2NaOsO_6$ ($0.60\mu_B$) [33]. Although an ideal $J = 3/2$ state would yield an effective magnetic moment of zero, hybridization between oxygen and transition metal orbitals and/or between the excited $J = 1/2$ state and the ground $J = 3/2$ state leads to deviations, resulting in a finite $\mu_{eff}$, as experimentally observed [34] [35]. Resonant inelastic x-ray scattering experiments confirm that the $J = 3/2$ state persists in double perovskites with $Re^{6+}$ ions, even with distortions in the $ReO_6$ octahedra [8]. While the $ReO_6$ octahedra in $Ba_2(Cd_{1-x}Ca_x)ReO_6$ ($x = 0$–1) elongate slightly below the cubic-to-tetragonal structural transition at $T_s$, it is reasonable to assume that the $J = 3/2$ state is preserved across the solid solutions, independent of composition.

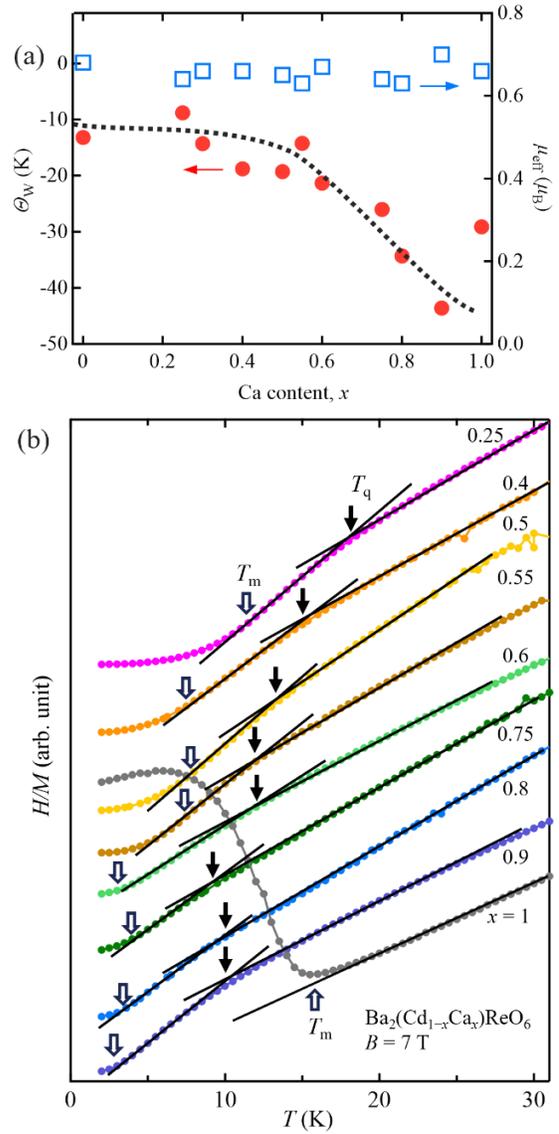

FIG. 3. (a) Compositional dependence of effective magnetic moments (blue open squares, right axis) and Weiss temperatures (red filled circles, left axis) of $Ba_2(Cd_{1-x}Ca_x)ReO_6$ ($x = 0$–1) derived from Curie-Weiss fits between 100 and 350 K. (b) Temperature dependence of the inverse susceptibilities below 30 K under a 7 T applied magnetic field. The quadrupole order transition temperature ($T_q$, filled arrows) and the magnetic transition temperatures ($T_m$, open arrows) are estimated from kinks in the inverse susceptibilities. The broken curve in (a) and black solid lines in (b) are provided as visual guides.

Negative Weiss temperatures across all compositions suggest dominant antiferromagnetic interactions. For $x = 0$–0.55, the observed cAFM order aligns with these dominant interactions despite the ferromagnetic behavior. Similarly, the dominant antiferromagnetic interactions are consistent with the collinear AFM phases at $x \geq 0.6$. For $x = 0$–0.55, the Weiss temperature's absolute value (10−20 K) approximately corresponds to $T_m$, while for $x \geq 0.6$, Weiss temperatures (~50 K) deviate significantly from $T_m$ (~3 K). The suppression of the magnetic transition temperature in these systems can be attributed to frustration caused by the geometrically frustrated FCC lattice arrangement of the transition metals in double perovskites.

### C. Quadrupole order

The temperature dependence of the inverse susceptibility at 7 T reveals a kink below 30 K for the solid solutions $Ba_2(Cd_{1-x}Ca_x)ReO_6$ ($x$ = 0.25–0.9) [Fig. 3(b)]. Between this kink and $T_m$, the inverse susceptibility exhibits a linear temperature dependence, indicative of a paramagnetic state. A similar kink is observed in $Ba_2CdReO_6$ ($x$ = 0), attributed to a quadrupolar ordering transition [23]. Theoretical studies have predicted such kinks, associating them with quadrupole order transitions and a Curie-Weiss region with a different slope below the kink [36]. It is plausible that the observed kinks in the Ca-substituted compounds also originate from quadrupole order transitions. In contrast, no such kink is detected in $Ba_2CaReO_6$ ($x$ = 1) across the measured temperature range.

The kink temperature, $T_q$, decreases systematically with increasing $x$, from 25 K in $Ba_2CdReO_6$ ($x$ = 0). Notably, there is no significant change in $T_q$ at $x$ = 0.6, where the magnetic ground state transitions. Kinks are observed for all solid solutions with $x \leq 0.9$, suggesting quadrupole ordering persists up to compositions close to $Ba_2CaReO_6$ ($x$ = 1).

### D. Electronic phase diagram

The experimentally determined electronic phase diagram for $Ba_2(Cd_{1-x}Ca_x)ReO_6$ ($x$ = 0–1) is shown in Fig. 4. For $x$ = 0–0.55 ($a$ = 8.323–8.355 Å), the magnetic transition temperature gradually decreases with lattice expansion, reaching $T_m \sim 7$ K. Beyond this, the magnetic state transitions from cAFM to collinear AFM. For $x$ = 0.6–0.9 ($a$ = 8.356–8.366 Å), $T_m$ stabilizes around 3 K, independent of the lattice constant. At $x$ = 1, a collinear AFM is observed at $T_m$ = 16 K, consistent with prior studies [25] [26].

The $T_q$ values decrease systematically with increasing lattice constant, and quadrupolar order is observed up to $x$ = 0.9. However, it vanishes at $x$ = 1, indicating that quadrupole order disappears between $x$ = 0.9 and 1. The electronic phase diagram reveals a previously unreported phase in double perovskites containing $Re^{6+}$ ions, characterized by intermediate quadrupole order coexisting with a collinear AFM ground state.

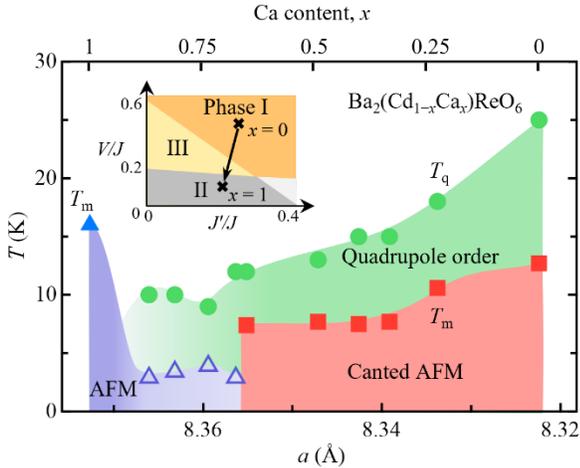

FIG. 4. Electronic phase diagram of $Ba_2(Cd_{1-x}Ca_x)ReO_6$ ($x$ = 0–1). The quadrupole ordering transition temperature ($T_q$, green filled circles) and magnetic transition temperature ($T_m$, red filled squares for cAFM and blue triangles for collinear AFM) are plotted against Ca concentration ($x$, upper axis) and room-temperature lattice constants ($a$, lower axis). The inset presents a schematic electronic phase diagram as a function of $J'/J$ (horizontal axis) and $V/J$ (vertical axis). In phase I, the system transitions from a normal state to a quadrupolar-ordered phase, and subsequently to a cAFM phase upon cooling. In Region II, a single transition to a collinear AFM state occurs. In Region III, the system exhibits an intermediate-temperature quadrupolar phase with a collinear AFM ground state. Black crosses indicate the approximate positions of $x$ = 0 and $x$ = 1, and the black arrow represents the trajectory of Ca substitution.

## IV. DISCUSSION

By comparing the experimentally obtained electronic phase diagram in Fig. 4 with that theoretically proposed one, we can identify the key electronic interactions governing the electronic phase that are modified by chemical substitution. The electronic phase diagram obtained in this study shares several features with that proposed by Chen et al., using mean-field theory for spin-orbit-entangled electrons with a $J$ = 3/2 state [4].

In the theoretical phase diagram, the electric quadrupole interaction ($V$) is considered alongside the nearest-neighbor antiferromagnetic ($J$) and ferromagnetic ($J'$) exchange interactions. Previous studies have suggested that double perovskites occupy a region where $J'/J$ and $V/J$ are relatively small.

Within this region, three distinct electronic phases (I, II, and III) are predicted, as depicted in the inset of Fig. 4. When $J'/J$ and $V/J$ exceed the line connecting $V/J$ = 0.6 and $J'/J$ = 0.4, a cAFM ground state emerges with intermediate-temperature quadrupole order (phase I). Conversely, in the region where $J'/J$ and $V/J$ are smaller than the line connecting $V/J$ = 0.2 and $J'/J$ = 1, the system transitions directly to a collinear AFM state (phase II). Between these two phases lies a region (phase III) characterized by a collinear AFM magnetic ground state below an intermediate-temperature quadrupole order. Notably, phase II also exhibits simultaneous octupolar order with the collinear AFM order. Phases I, II, and III correspond to the electronic phases for $0 \leq x \leq 0.55$, $x$ = 1 ($Ba_2CaReO_6$), and $0.6 \leq x \leq 0.9$, respectively.

The correspondence between the experimental results and the theoretical phase diagram suggests that $Ba_2(Cd_{1-x}Ca_x)ReO_6$ ($x$ = 0–1) exists near the boundaries of the three phases. At the boundary between phases I and III, theory predicts a magnetic ground state that transitions smoothly from cAFM to collinear AFM order, with a continuous reduction in the cant angle, consistent with the observed decrease in net moment. This transition is likely driven by decreases in $J'/J$ and $V/J$.

The compositional dependence of the magnetic interaction $J'/J$ can be inferred from the compositional variation of the Weiss temperature. Since the Weiss temperature, which represents the sum of ferromagnetic and antiferromagnetic interactions, is negative for all compositions, $J$ consistently exceeds $J'$ ($J'/J \leq 1$). The increasing absolute value of the Weiss temperature with increasing $x$ indicates a relative reduction in the ferromagnetic interaction ($J'/J$), consistent with the theoretical phase diagram. As the lattice constant at room temperature increases linearly with $x$, the Re–Re distances are expected to follow the same trend. However, the Weiss temperature exhibits non-linear behavior, decreasing significantly beyond $x = 0.6$. The exchange paths for interactions $J$ and $J'$ are Re–O–O–Re [Fig. 1(b)], and the magnetic interaction arises through a virtual transfer through the intermediate oxygen $p$ orbitals [16]. Despite the lattice constants varying linearly with Ca content, $x$, the nonlinear change in magnetic interactions may indicate that the degree of hybridization between the Re $d$ orbitals and the oxygen $p$ orbitals also changes nonlinearly.

The electric quadrupole interaction ($V$) is given by $V = 9\sqrt{2}Q^2/a^5$ [16], where $Q$ and $a$ represent the magnitude of the electric quadrupole and the lattice constant, respectively. $V$ is expected to decrease with lattice expansion due to increases in $a$ and reductions in $Q$. Typically, $Q$ is dominated by the charge at oxygen sites. Increased Re–O distance due to lattice expansion weakens the hybridization between Re $5d$ and O $2p$ orbitals, reducing $Q$. For $Ba_2CdReO_6$ ($x = 0$), the average Re–O distance is 1.918(8) Å [27], compared to 1.954(16) Å [25] for $Ba_2CaReO_6$ ($x = 1$).

The decrease in $V$ is supported by changes in $T_q$, which is expressed as $T_q = \frac{43V + 18J' - 3J}{18}$ [16]. In the region $x \leq 0.55$, where the Weiss temperature changes minimally, $J/J'$ remains relatively constant. The monotonic decline in $T_q$ with increasing lattice constant confirms the reduction in $V$. Thus, lattice expansion from substituting Ca for $Ba_2CdReO_6$ reduces both $J'/J$ and $V/J$, driving the transitions among the three electronic phases.

Finally, the electronic phase diagram of $Ba_2(Cd_{1-x}Ca_x)ReO_6$ is compared with that of $Ba_2MgReO_6$ under applied pressure. In $Ba_2MgReO_6$, $T_q$ remains nearly constant up to 5 GPa, but is abruptly suppressed near 6 GPa [17]. The cAFM ground state at ambient pressure transitions to collinear AFM order alongside suppression of quadrupolar order. Unlike Ca substitution, which causes lattice expansion, applied pressure is expected to increase $V/J$. Materials with smaller lattice constants than $Ba_2MgReO_6$ tend to exhibit more negative Weiss temperatures, suggesting $J'/J$ decreases under pressure. Consequently, the simultaneous increase in $V/J$ and decrease in $J'/J$ may explain the constancy of $T_q$ [17]. However, recent DFT studies suggest that the cAFM phase and high-temperature quadrupole-ordered phase are resilient to isotropic pressure [18]. The experimentally observed phase transitions imply that anisotropic pressure plays a critical role. This highlights the potential for greater control of interactions via anisotropic pressure, opening new avenues for future research.

## V. CONCLUSIONS

In conclusion, we synthesized and investigated the electronic properties of the solid solutions of $5d^1$ double perovskites $Ba_2CaReO_6$ and $Ba_2CdReO_6$. The compound $Ba_2(Cd_{1-x}Ca_x)ReO_6$ exhibits a change in the multipolar ground state upon chemical substitution. Across all compositions, the effective magnetic moment is suppressed, confirming the presence of a spin-orbit-entangled $J = 3/2$ state. The saturation magnetic moment decreases with increasing Ca content, suggesting a reduction in the canting angle of the cAFM order. Quadrupolar order was observed for all compositions except $x = 1$, with $T_q$ decreasing as the lattice constant increased. Theoretical correspondence suggests that lattice expansion, driven by chemical pressure, reduces the $V/J$ and $J'/J$ ratios, thereby suppressing quadrupolar and cAFM ordering.

The strong agreement between the electronic phase diagram obtained in this study and that predicted by mean-field theory underscores the critical role of the three interactions used in the mean-field approximation in determining electronic phases. Furthermore, this study demonstrates that these interactions can be effectively tuned through chemical pressure. These findings are expected to serve as a valuable guideline for the future exploration of novel multipole orders in spin-orbit-entangled $d$-electron systems.

## ACKNOWLEDGEMENTS

The authors thank Z. Hiroi, T. Muroi, and N. Iwahara for fruitful discussions. This paper was partly conducted under the Visiting Researcher's Program of the Institute for Solid State Physics, the University of Tokyo. This paper was partly supported by the Japan Society for the Promotion of Science KAKENHI Grants No. JP20H01858, No. JP23H04860, and No. JP 24H01187 and the JSPS Bilateral Open Partnership Joint Research ProjectsJPJSBP120239915.

## REFERENCES

[1] A. Abragam and B. Bleaney, *Electron Paramagnetic Resonance of Transition Ions* (Oxford University Press, Oxford, New York, 2012).

[2] G. Jackeli and G. Khaliullin, Mott Insulators in the Strong Spin-Orbit Coupling Limit: From Heisenberg to a Quantum Compass and Kitaev Models, Phys Rev Lett **102**, 017205 (2009).

[3] J. G. Rau, E. K.-H. Lee, and H.-Y. Kee, Spin-Orbit Physics Giving Rise to Novel Phases in Correlated Systems: Iridates and Related Materials, Annu Rev Condens Matter Phys **7**, 195 (2016).

[4] W. Witczak-Krempa, G. Chen, Y. B. Kim, and L. Balents, Correlated quantum phenomena in the strong spin-orbit regime, Annu Rev Condens Matter Phys **5**, (2014).

[5] T. Takayama, J. Chaloupka, A. Smerald, G. Khaliullin, and H. Takagi, Spin–Orbit-Entangled Electronic Phases in $4d$ and $5d$ Transition-Metal Compounds, J Physical Soc Japan **90**, 062001 (2021).

[6] C. A. Marjerrison et al., Cubic $Re^{6+}$ ($5d^1$) Double Perovskites, $Ba_2MgReO_6$, $Ba_2ZnReO_6$, and $Ba_2Y_{2/3}ReO_6$: Magnetism, Heat Capacity, μSR, and Neutron Scattering Studies and Comparison with Theory, Inorg Chem **55**, 10701 (2016).

[7] D. Hirai and Z. Hiroi, Successive Symmetry Breaking in a $J_{eff} = 3/2$ Quartet in the Spin–Orbit Coupled Insulator $Ba_2MgReO_6$, J Physical Soc Japan **88**, 064712 (2019).

[8] F. I. Frontini, G. H. J. Johnstone, N. Iwahara, P. Bhattacharyya, N. A. Bogdanov, L. Hozoi, M. H. Upton, D. M. Casa, D. Hirai, and Y.-J. Kim, Spin-Orbit-Lattice Entangled State in $A_2MgReO_6$ (A = Ca, Sr, Ba) Revealed by Resonant Inelastic X-Ray Scattering, Phys Rev Lett **133**, 36501 (2024).

[9] D. Hirai, H. Sagayama, S. Gao, H. Ohsumi, G. Chen, T. Arima, and Z. Hiroi, Detection of multipolar orders in the spin-orbit-coupled $5d$ Mott insulator $Ba_2MgReO_6$, Phys Rev Res **2**, 022063 (2020).

[10] D. D. Maharaj, G. Sala, M. B. Stone, E. Kermarrec, C. Ritter, F. Fauth, C. A. Marjerrison, J. E. Greedan, A. Paramekanti, and B. D. Gaulin, Octupolar versus Néel Order in Cubic $5d^2$ Double Perovskites, Phys Rev Lett **124**, 087206 (2020).

[11] A. Paramekanti, D. D. Maharaj, and B. D. Gaulin, Octupolar order in $d$-orbital Mott insulators, Phys Rev B **101**, 054439 (2020).

[12] S. Vasala and M. Karppinen, $A_2B'B''O_6$ perovskites: A review, Progress in Solid State Chemistry **43**, 1 (2015).

[13] H. Kusunose, Description of Multipole in $f$-Electron Systems, J Physical Soc Japan **77**, 064710 (2008).

[14] Y. Kuramoto, H. Kusunose, and A. Kiss, Multipole Orders and Fluctuations in Strongly Correlated Electron Systems, J Physical Soc Japan **78**, 072001 (2009).

[15] P. Santini, S. Carretta, G. Amoretti, R. Caciuffo, N. Magnani, and G. H. Lander, Multipolar interactions in $f$-electron systems: The paradigm of actinide dioxides, Rev Mod Phys **81**, 807 (2009).

[16] G. Chen, R. Pereira, and L. Balents, Exotic phases induced by strong spin-orbit coupling in ordered double perovskites, Phys Rev B **82**, 174440 (2010).

[17] H. Arima, Y. Oshita, D. Hirai, Z. Hiroi, and K. Matsubayashi, Interplay between Quadrupolar and Magnetic Interactions in $5d$ Double Perovskite $Ba_2MgReO_6$ under Pressure, J Physical Soc Japan **91**, 013702 (2022).

[18] D. Fiore Mosca, C. Franchini, and L. V Pourovskii, Interplay of superexchange and vibronic effects in the hidden order of $Ba_2MgReO_6$ from first principles, Phys Rev B **110**, L201101 (2024).

[19] R. D. Shannon, Revised effective ionic radii and systematic studies of interatomic distances in halides and chalcogenides, Acta Crystallographica Section A **32**, 751 (1976).

[20] D. Hirai, A. Koda, A. Matsuo, K. Kindo, T. Yajima, and Z. Hiroi, *Muon Spin Rotation, High-Field Magnetization, and Structural Study on a Spin–Orbit-Entangled Mott Insulator $Ba_2MgReO_6$*, in *JPS Conference Proceedings* (Physical Society of Japan, 2020), p. 011143.

[21] J. Longo and R. Ward, Magnetic Compounds of Hexavalent Rhenium with the Perovskite-type Structure, J Am Chem Soc **83**, 2816 (1961).

[22] A. W. Sleight, J. Longo, and R. Ward, Compounds of Osmium and Rhenium with the Ordered Perovskite Structure, Inorg Chem **1**, 245 (1962).

[23] D. Hirai and Z. Hiroi, Possible quadrupole order in tetragonal $Ba_2CdReO_6$ and chemical trend in the ground states of $5d^1$ double perovskites, Journal of Physics Condensed Matter **33**, 135603 (2021).

[24] B. L. Chamberland and G. Levasseur, Rhenium oxides having an ordered or related perovskite-type structure, Mater Res Bull **14**, 401 (1979).

[25] K. Yamamura, M. Wakeshima, and Y. Hinatsu, Structural phase transition and magnetic properties of double perovskites $Ba_2CaMO_6$ (M = W, Re, Os), J Solid State Chem **179**, 605 (2006).

[26] H. Ishikawa, D. Hirai, A. Ikeda, M. Gen, T. Yajima, A. Matsuo, Y. H. Matsuda, Z. Hiroi, and K. Kindo, Phase transition in the $5d^1$ double perovskite $Ba_2CaReO_6$ induced by high magnetic field, Phys Rev B **104**, 174422 (2021).

[27] V. da Cruz Pinha Barbosa et al., The Impact of Structural Distortions on the Magnetism of Double Perovskites Containing $5d^1$ Transition-Metal Ions, Chemistry of Materials **34**, 1098 (2022).

[28] I. Živković et al., Dynamic Jahn-Teller effect in the strong spin-orbit coupling regime, Nat Commun **15**, 8587 (2024).

[29] J.-R. Soh et al., Spectroscopic signatures and origin of hidden order in $Ba_2MgReO_6$, Nat Commun **15**, 10383 (2024).

[30] V. da Cruz Pinha Barbosa et al., Exploring the Links between Structural Distortions, Orbital Ordering, and Multipolar Magnetic Ordering in Double Perovskites Containing Re(VI) and Os(VII), Chemistry of Materials (2024).

[31] F. R. Kroeger and C. A. Swenson, Absolute linear thermal-expansion measurements on copper and aluminum from 5 to 320 K, J Appl Phys **48**, 853 (1977).

[32] V. M. Goldschmidt, Die Gesetze der Krystallochemie, Naturwissenschaften **14**, 477 (1926).

[33] A. S. Erickson, S. Misra, G. J. Miller, R. R. Gupta, Z. Schlesinger, W. A. Harrison, J. M. Kim, and I. R. Fisher, Ferromagnetism in the mott insulator $Ba_2NaOsO_6$, Phys Rev Lett **99**, 016404 (2007).

[34] K. H. Ahn, K. Pajskr, K. W. Lee, and J. Kuneš, Calculated $g$-factors of $5d$ double perovskites $Ba_2NaOsO_6$ and $Ba_2YOsO_6$, Phys Rev B **95**, 064416 (2017).

[35] H. Kubo, T. Ishitobi, and K. Hattori, Electronic origin of ferroic quadrupole moment under antiferroic quadrupole order and finite magnetic moment in $J_{eff}$ = 3/2 systems, Phys Rev B **107**, 235134 (2023).

[36] C. Svoboda, W. Zhang, M. Randeria, and N. Trivedi, Orbital order drives magnetic order in $5d^1$ and $5d^2$ double perovskite Mott insulators, Phys Rev B **104**, 24437 (2021).